\documentclass[prb,preprint]{revtex4-1} 

\usepackage{xcolor}
\usepackage{amsmath}  
\usepackage{amsfonts} 
\usepackage{graphicx} 
\usepackage[utf8]{inputenc}
\usepackage{gensymb}
\usepackage{graphicx}
\usepackage{color, colortbl}
\definecolor{LightCyan}{rgb}{0.88,1,1}
\usepackage{multirow}

\begin{document}

\title{Differences in the attitudes and beliefs about science of
  students in the physics-mathematics and life sciences areas and
  their impact on teaching 
  }

\author{Álvaro Suárez, Daniel Baccino}
\affiliation{Consejo de Formación en Educación, Instituto de Profesores Artigas, Montevideo, Uruguay}
\author{Martin Monteiro}
\affiliation{Universidad ORT Uruguay,Montevideo Uruguay}
\author{Arturo C. Marti}
\email{marti@fisica.edu.uy}

\affiliation{Instituto de F\'{i}sica, Universidad de la Rep\'{u}blica, Uruguay}

\begin{abstract}
For this study, we compared the attitudes and beliefs
about science of physical science (physics and mathematics) and life
science (biochemistry and biology) students at the beginning of their
university degrees using the CLASS (Colorado Learning Attitudes about
Science Survey) tool. It is worth noting that both groups of students
received similar physics courses during their high-school
education. Through a detailed analysis of the different categories of
the test, we examined the differences in performance in each of the
areas that make up the questionnaire. Among other aspects, we found
that a considerable percentage of life science students (higher than
that of physical science students) adopted a novice type of behavior
in problem solving. Finally, we discussed the possible causes of the
differences found and their implications for teaching.
\end{abstract}

\maketitle

\section{Introduction}
\label{sec:int}

It has long been recognized in the physics-teaching community that, in
addition to the specific contents, certain characteristics of the
students, such as their previous ideas, their personal history, the
way they interact with their peers and teachers, and their
expectations, determine the transformation that occurs as they move
through the educational system \cite{docktor2014synthesis}. In
particular, the set of ideas, assumptions and previous conceptions
about science—more specifically, its evolution, its methods, its
validation or refutation—are encompassed in the concept of
epistemological beliefs. These beliefs, which are generally not
explicit, have a strong impact on teaching and learning
\cite{redish1998student}. Therefore, it is relevant to understand the
attitudes and beliefs of our students and how educational
institutions, their teachers and teaching materials directly or
indirectly affect those attitudes and beliefs in the learning and
teaching of physics.

In the field of Physics Education Research (PER) a set of tools have
been developed for the systematic assessment of knowledge, focusing on
aspects related to the discipline as well as on other less objective
aspects regarding the students' attitudes and beliefs
\cite{redish1998student,adams2006new,wilcox2016students}. While the
assessment of specific knowledge is mostly based on multiple-choice
questionnaires, the tools oriented to attitudes and beliefs usually
ask students to indicate their degree of agreement or disagreement
(Likert scale) with different statements reflecting the opinion of
"experts" (usually professional physicists). The questionnaires result
from a design and validation process that includes repeated
interactions with experts \cite{redish1998student,adams2006new}. The
differences between the students' responses and the experts' responses
constitute the raw material for analyzing the epistemological status
of the group of students. Specifically, to quantify the impact of
certain courses, interventions or teaching approaches, it is applied
twice, once at the beginning of the course (pre-test) and once after
the intervention (post-test). The comparison of results reflects
changes in attitudes and beliefs as a function of demographics,
methodologies and teaching strategies.

Among the standardized tests for the evaluation of epistemological
beliefs, we highlight the MPEX (Maryland Physics Expectations Survey)
\cite{redish1998student}, which aims to survey the expectations of
students in relation to physics, the CLASS (Colorado Learning
Attitudes about Science Survey) \cite{adams2006new},
aimed at assessing the attitudes of students toward learning
physics, how they think physics relates to everyday life, and their
opinion about the discipline, and the E-CLASS (Colorado Learning
Attitudes about Science Survey for Experimental Physics)
\cite{wilcox2016students}, which follows the same approach as the
CLASS but aimed at the experimental aspects of physics.

Research on the attitudes and beliefs of students has provided
valuable insights into the discipline
\cite{madsen2015physics}. Several studies have shown that some
epistemological attitudes negatively affect learning
\cite{perkins2010becomes,milner2011attitudes}. For example, a student
who sees physics as a set of unconnected facts and formulas will study
differently than one who sees it as a network of interconnected
concepts \cite{redish1998student}. Another valuable input shows that
there is a positive correlation of academic performance with the CLASS
pre-test and MPEX scores
\cite{perkins2010becomes,cahill2018understanding}. In addition,
multiple quantitative studies have been conducted in recent years on
the changes that occur in students' attitudes and beliefs as a
function of their previous training, the types of courses, and the
teaching strategies used, as well as on the relationship between the
pre-test and academic performance, among other variables
\cite{madsen2015physics}.

Also valuable is the work comparing the performance of different
groups in questionnaires regarding attitudes and beliefs about
science. An extensive meta-analysis shows that students who major in
physics perform better than those who major in engineering or other
sciences \cite{madsen2015physics}. This study also suggests that these
aspects develop in the early stages of education, although it leaves
several questions to be answered as to why this phenomenon
occurs. Another longitudinal study \cite{perkins2010becomes}, followed
the journey of a group of students throughout their university careers
and showed that those who were determined to study physics at the
beginning of the first year of college performed better than the
average of their peers.

Although the application of the CLASS and other similar instruments to
assess epistemological attitudes has spread to many parts of the
world, Latin American countries are a step behind, as very little
research has been conducted in the area
\cite{tecpan2015actitudes,injoque2018actitudes,santana2018ganancia}. In
the particular case of Uruguay, we inquired about the epistemological
beliefs of teachers and prospective high school physics teachers in
Uruguay \cite{suarez2021cu}. The results showed trends shared by both
groups, especially in the categories related to personal effort,
interest and connection with the real world. However, there were
notable differences in other categories, particularly those related to
conceptual understanding, confidence and sophistication in problem
solving.

There are many open questions regarding the attitudes and beliefs of
students, especially those related to the comparison between different
groups, the causes of possible differences and their impact on
learning. In this work, we compare the attitudes and beliefs about
science of students of physical sciences (physics and mathematics)
with others of life sciences (biochemistry and biology) at the
beginning of their university degrees using the CLASS tool. Our study
was conducted in the first year of the degrees, at a very early stage
when the students’ curricular training in physics comes mainly from
high-school courses. Through a detailed analysis of the different
categories into which the CLASS tool is divided, we analyze the
differences in performance in each of the areas and discuss their
possible causes and implications for teaching. In the following
section we present the research methodology, while in section 3 we
present the main results. Finally, the discussion and final
considerations are presented in the last two sections.

\section{Research methodology}
\label{sec:meth}

This research was carried out with university students from the School
of Sciences of Universidad de la República (Montevideo, Uruguay) who
were taking General Physics I in the first semester of the
undergraduate courses in Physics and Mathematics and Biology and
Biochemistry. It is relevant to point out that in Uruguay, primary,
secondary and pre-university schools are characterized by a common
curriculum framework for all the educational institutions in the
country. In the last two years of high school, students are able to
choose a major area of study (scientific, humanistic, biological or
artistic). Those who wish to pursue university studies in scientific
areas choose a path with more hours devoted to physics and mathematics
courses, while those who plan to pursue degrees in life sciences, such
as biology or biochemistry, have their hours split between
mathematics, physics, chemistry and biology courses. It is important
to point out that in both orientations, physics courses have the same
weekly load and similar contents that address general topics of
mechanics (kinematics and point dynamics, principles of conservation
and energy) as well as waves and electromagnetism. The usual
bibliography includes algebra-based “College Physics” textbooks such
as those widely used worldwide
\cite{serway2014college,wilson2019college}. Upon completing high
school, students can opt for different university degrees. In this
work we focus on two sets of recently admitted students in the first
year of the aforementioned School of Science. One set is comprised by
those who pursue bachelor’s degrees in Physics and Mathematics
(hereinafter, “physical sciences” or “PhS”) and the other by those who
pursue bachelor's degrees in Biology and Biochemistry (hereinafter,
“life sciences” or “LS”).

With this objective in mind, we asked all the students to answer, by
means of an electronic form, their degree of agreement or disagreement
with the 42 statements of the CLASS test. To compare the students'
responses with those of the experts, the original 5-level scale is
reduced to a three-level scale, grouping the options "agree" and
"completely agree" on the one hand, and "disagree" and completely
disagree" on the other, leaving "neutral" as the midpoint. Of a total
of 42 statements, 27 of them are grouped into 8 categories, while the
remaining 15 are not categorized. There is no agreement among the
experts on some of the latter, while one in particular is used to rule
out inconsistent responses. For each category, and for the set of
questions where there is agreement among the experts (36 of the 42
statements), we calculated the percentage of student responses that
agree with that of the experts (favorable responses). Table 1 shows
the number of responses obtained for each group of students.

\begin{table}
\begin{center}
 \begin{tabular}{|c|c|c|c|}
\hline
Bachelor’s degree     & Women  & Men & Total  \\  \hline  \hline 
Physical sciences& 34 & 45 & 79 \\  \hline
Life sciences  & 51& 18& 69 \\   \hline
\end{tabular} 
\end{center}
\caption{\label{tab:table-name}Number of responses recorded in 
the different areas of study classified by self-reported gender.}
\end{table}

In order to study the differences between the two samples of results,
we used the nonparametric Mann-Whitney U test, which is the
nonparametric version of Student's t-test \cite{mann1947test}, to test
if the differences observed were statistically significant. For this
purpose, we started from the null hypothesis that the samples come
from the same population, rejecting it for a probability value of
p<0.05. To quantify the possible differences between the two groups,
we calculated the effect size using Cohen's d. Typically, a large
effect is considered for d values of 0.8, while medium corresponds to
0.5 and small to 0.2 (Cohen, 2013).

\section{Results}
\label{sec:res}

In this section we show the results obtained from the CLASS
questionnaire, differentiating the groups of students of life sciences
(LS) from those of physical sciences (PhS). In Table 2, we indicate,
for each category, the mean values of the LS and PhS groups and the
difference between them (columns 2, 3 and 4, respectively), with their
corresponding standard errors. In the fifth column we tabulate the
effect size for the difference between the means, and in the sixth
column we indicate the probability given for the result of applying
the nonparametric Mann-Whitney U test.

We highlight some results that can be deduced from Table 2. First,
when considering the overall result of CLASS (All categories), we
found a significant difference, with a relatively large effect size
(0.8), in the degree of agreement with the experts between both groups
of students. By applying the Mann-Whitney U statistical test to the
samples of each category grouped by area (LS and PhS), we found that
the null hypothesis (equality of mean values) can be rejected in all
categories, except in Real-World Connection. The largest effect size
values correspond to the following categories: Problem Solving -
Sophistication (1.1), Applied Conceptual understanding (0.8), Problem
Solving - General (0.7) and Problem Solving - Confidence (0.7). Both
groups of students show a significant difference in their agreements
with experts in the vast majority of the CLASS categories. We cannot
rule out the hypothesis of equality between the central tendencies of
the samples in the Real-World Connection category at the 95\%
confidence level (highlighted row in Table 2). This is consistent with
the fact that the value of the effect size of the difference is the
smallest of all those obtained (0.4). The LS and PhS students surveyed
start their undergraduate training with similar conceptions in this
CLASS category.

\begin{table}
\begin{center}
 \begin{tabular}{|c|c|c|c|c|c|} \hline
-& LS& PhS& LS-PhS diff.& Effect size &
U test: p \\  \hline  \hline 
All categories & 49.8(2.2) & 64.1(3.2) & 14.3 (3.7) & 0,8 & 0,0001 \\  \hline
Personal Interest & 51.9 (3.2)& 73.7 (4.1)& 21.8 (5.3)& 0,8 & 0,0001 \\  \hline \rowcolor{LightCyan}
Real-World Connection & 58.3 (3.8)& 70.4 (4.9)&12.0 (6.3)& 0,4& 0,0542 \\  \hline
Problem Solving (PS) - General & 47.8 (2.8) & 65.1 (4.1)& 17.3 (4.8) & 0,7 & 0,0004 \\  \hline
PS Confidence& 41.7 (3.3)& 61.8 (5.2)& 20.2 (5.9)& 0,7& 0,0009  \\  \hline
PS Sophistication& 27.5 (2.6)&54.4 (4.3) &26.8 (4.8) & 1,1& 0,0001   \\  \hline 
Sense Making/Effort & 65.2 (2.6)& 73.7 (3.9)& 8.5 (4.5)& 0,4& 0,0226   \\  \hline 
Conceptual understanding& 49.8(2.9)& 62.7 (3.6)& 13.0 (4.7)& 0,5& 0,0062 \\  \hline
Applied Conceptual understanding& 32.3 (2.4)& 50.4 (4.2)& 18.1 (4.5)& 0,8& 0,0004   \\  \hline
\end{tabular} 
\end{center}
\caption{\label{tab:t2} Overall results of the CLASS questionnaire discriminated by group. The columns indicate mean value and difference between the LS and PhS groups with their standard errors, effect size of the difference and probability according to the Mann-Whitney U test. The highlighted row corresponds to the case in which we cannot rule out the hypothesis of equality between the central tendencies of the samples.
}
\end{table}

In the Sense Making/Effort category, we identified a "borderline" or
less noticeable situation than in the other categories in which
significant differences were recorded. In that category, we observed
the smallest difference in means, an effect size of the same value as
in the Real-World Connection category, and a probability of p=0.02 for
the U test (of the same order as our 0.05 limit). In this category,
the difference between the samples (which exists according to the
criterion adopted) is not as clear as in the other categories where
there is a significant difference.

Although in each of the above categories we can find statements whose
results differ significantly between the two groups of students, we
compare the results of two particular categories: Problem Solving -
Sophistication and Applied Conceptual Understanding, since these are
the categories with the largest effect size. Table 3 shows the
statements of both categories for each group of students, specifying
the percentages of the responses, classified as agree, neutral and
disagree. The underlined percentages allow us to identify whether the
experts agree or disagree with each of the statements. For example,
statement 5 falls into both categories, with the experts disagreeing
with it. Of the set of statements, in statements 6 and 21 the
difference between the responses of the students in both groups is
less than the standard deviation of the responses, so it is not of
particular interest to analyze it. In the following section we discuss
the results and their possible implications for the classroom and
student learning.

\begin{table}[ht]
\tiny
\begin{center}
\begin{tabular}{p{0.35\textwidth}p{0.18\textwidth}p{0.07\textwidth}p{0.05\textwidth}p{0.05\textwidth}p{0.05\textwidth}}
    \hline
    Statement & Category & Area & Agree & Neutral & Disagree\\
    \hline
     \multirow{2}{=}{1. A significant problem in learning physics is being able to memorize all the information I need to know } & \multirow{2}{=}{Applied Conceptual understanding} & Life sciences & 41\% & 28\% & 32\%\\
     & & Physical sciences & 26\% & 26\% & 47\%\\
    \hline
    \multirow{2}{=}{5. After I study a topic in physics and feel that I understand it, I have difficulty solving problems on the same topic } & \multirow{2}{=}{Applied Conceptual understanding and PS Sophistication} & Life sciences & 77\% & 13\% & 10\%\\
     & & Physical sciences & 42\% & 32\% & 26\%\\
     \hline
     \multirow{2}{=}{6. Knowledge in physics consists of many disconnected topics. } & \multirow{2}{=}{Applied Conceptual understanding} & Life sciences & 9\% & 14\% & 77\%\\
     & & Physical sciences & 5\% & 11\% & 84\%\\
     \hline
     \multirow{2}{=}{8. When I solve a physics problem, I locate an equation that uses the variables given in the problem and plug in the values } & \multirow{2}{=}{Applied Conceptual understanding} & Life sciences & 80\% & 19\% & 1\%\\
     & & Physical sciences & 39\% & 37\% & 24\%\\
     \hline
     \multirow{2}{=}{21. If I don't remember a particular equation needed to solve a problem on an exam, there's nothing much I can do (legally!) to come up with it. } & \multirow{2}{=}{Applied Conceptual understanding and PS Sophistication} & Life sciences & 28\% & 17\% & 55\%\\
     & & Physical sciences & 24\% & 11\% & 66\%\\
     \hline
     \multirow{2}{=}{22. If I want to apply a method used for solving one physics problem to another problem, the problems must involve very similar situations. } & \multirow{2}{=}{Applied Conceptual understanding and PS Sophistication} & Life sciences & 45\% & 35\% & 20\%\\
     & & Physical sciences & 34\% & 24\% & 20\%\\
     \hline
     \multirow{2}{=}{25. I enjoy solving physics problems. } & \multirow{2}{=}{PS Sophistication} & Life sciences & 20\% & 33\% & 46\%\\
     & & Physical sciences & 79\% & 13\% & 8\%\\
     \hline
     \multirow{2}{=}{34. I can usually figure out a way to solve physics problems. } & \multirow{2}{=}{PS Sophistication} & Life sciences & 29\% & 41\% & 30\%\\
     & & Physical sciences & 50\% & 34\% & 16\%\\
     \hline
     \multirow{2}{=}{40. If I get stuck on a physics problem, there is no chance I'll figure it out on my own. } & \multirow{2}{=}{Applied Conceptual understanding and PS Sophistication} & Life sciences & 35\% & 35\% & 30\%\\
     & & Physical sciences & 8\% & 29\% & 63\%\\
     \hline
     \multicolumn{6}{p{0.85\textwidth}}{Table 3. Results of selected statements of the CLASS, classified by area of study (life sciences and physical sciences). Statements 5, 21, 22 and 40 belong to both the Applied Conceptual understanding and the PS Sophistication categories.}\\
    \hline
\end{tabular}
\end{center}
\label{tab:multicol}
\end{table}


\section{Discussion}
\label{sec:dis}

From the results presented in Table 3 for statements of all
categories, we note that there is a significant difference in
performance on the CLASS between students pursuing degrees in physical
sciences and those pursuing degrees in life sciences, with a large
effect size \cite{cohen2013statistical}. This result, in line with
others reported in the literature
\cite{gire2009characterizing,perkins2010becomes,bates2011attitudes},
supports the hypothesis that attitudes and beliefs develop during
secondary education, which becomes more relevant in the context of the
Uruguayan educational system, since students in both groups had a
similar number of hours devoted to physics courses in their previous
level of education, differing only in their training in mathematics
and biology. The roots of these differences could be very diverse. A
first hypothesis is that a better training in mathematics has a
positive impact on attitudes and beliefs. Another possibility is that
those students who have a taste for physical sciences from an early
age may have received a non-formal education (through books, articles
and dissemination videos) that results in a better performance in the
CLASS.

As for the other categories, Personal Interest, PS Sophistication and
Applied Conceptual understanding present a large effect size. This
result is not surprising, since students of physical science-oriented
degrees have a special interest in physics and mathematics, as
emphasized by \cite{adams2006new}. We clearly observe this in
statement 25, grouped in the PS Sophistication and Personal Interest
categories, which refers to enjoying physics problem solving. In the
case of the LS group of students, 20 \% agree with the experts,
whereas the level of agreement reaches 80\% among those studying PhS.
In relation to the large effect size for PS Sophistication and Applied
Conceptual understanding, we may hypothesize that students with a
biology-oriented background have not delved into sophisticated aspects
of problem solving nor into conceptual understanding in their physics
courses. This aspect could be linked to a lower degree of interest in
physics and perhaps to greater difficulties with the subject. These
two elements may foster negative attitudes towards physics and,
consequently, epistemological beliefs farther away from those of the
experts.  By focusing on the statements related to PS Sophistication
and Applied Conceptual understanding we can better understand how
different the attitudes and beliefs of PhS students may be from those
of LS students and their possible impact in the classroom. Let us
begin by analyzing statements 1 and 8, related to Applied Conceptual
understanding, on which the experts disagree. In statement 1, 41\% of
the LS students agree with the statement regarding the importance of
memorizing all the information in physics, while the percentage drops
to 26\% among PhS students. In statement 8, a large majority of the LS
students (80\%) agree with the methodology described for trying to
solve physics problems, dropping to 39\% in the other group of
students. From the results of statement 8, we can infer that the
majority of LS students attempt to solve problems using the "Plug and
Chug" strategy \cite{larkin1980expert,walsh2007phenomenographic},
which is a clear sign of a novice problem-solving strategy. This
aspect is consistent with the results of statement 1, which gives an
important role to memorization in physics learning.  Statements 5, 22
and 40 have the particularity that they belong to both categories and
that the experts disagree with them. These statements are closely
related to each other and to the way in which students learn, apply
concepts, develop metacognition and cope with physics problems. As in
the previously analyzed statements, the performance of LS students is
inferior to that of PhS students. The most alarming case is statement
5, where 77\% of LS students recognize that they have difficulties in
solving problems after considering that they have understood a
topic. This is a clear sign that they have not developed metacognition
habits and therefore are not able to realize that they do not truly
understand the concepts. In statement 40, which evaluates a student's
ability to think of different strategies to solve a problem when they
get stuck, it is inferred that one out of three LS students is not
able to develop new ways of solving, while this happens to only 8\% of
PhS students. This result is yet another sign that the "Plug and Chug"
strategy predominates in problem solving among LS students. Students
who have expert behaviors develop metacognition, are able to monitor
their learning, think borderline scenarios to evaluate solutions, and
try new strategies based on general principles; therefore, if they get
stuck on a problem, they are more likely to try to solve it using a
different approach. Statement 22 refers to the need for two problems
to be very similar in order to use the same solution method for
both. These results once again reinforce the idea that LS students, to
a greater extent than PhS students, fail to fully grasp the conceptual
aspects and basic principles of physics, and instead learn physics by
memorizing equations, which prevents them from approaching new
situations with the confidence to solve them.  Finally, in statement
34, which belongs to the PS Sophistication category and reflects
students' confidence in problem solving, only 29\% of LS students
believe they are able to figure out how to solve a physics problem,
compared to 50\% of PhS students. These results again suggest that LS
students find it difficult to adopt similar strategies as the experts
in problem solving, as well as perhaps a belief that they are not
capable of learning physics adequately.

\section{Final comments}
\label{sec:fin}

Our students' views on the nature of knowledge and learning work
either for or against quality science education by affecting the way
they learn and approach physics courses. For example, students who
view learning as being basically about memorizing information will
have different attitudes and strategies than those who view it as
being based on understanding \cite{elby2001helping}. In this sense,
one of the keys to improve learning is to promote appropriate
epistemological stances. Knowing the state of our students'
epistemological beliefs is crucial in order to design activities aimed
at improving them.  Our work is based on the results of proposing the
CLASS questionnaire to students of physical sciences and life
sciences. Comparing the results between both groups, we find that the
former enter university with epistemological attitudes much closer to
those of experts in the respective fields. Although this result has
been previously reported in the literature \cite{madsen2015physics},
our research differs in that the groups of students of physical
sciences and life sciences had a similar academic trajectory in terms
of their high school physics training. Although the question of the
possible origins of this difference is yet to be addressed, the
results of our study allow us to better understand the difficulties
presented by life sciences students when approaching physics courses.

From the analysis of the responses in the different categories, we
highlight that a significant percentage of life sciences students try
to solve physics problems following the strategy of "finding the right
equation and substituting", known as "Plug and Chug". This novice
strategy is not conducive to the development of metacognitive
skills. Being aware of this type of thinking allows us, as teachers,
to anticipate the problems of our students, to stand differently in
the classroom, and to develop actions aimed at changing this type of
thinking, which will result in a better-quality science education. In
this sense, it is important to keep in mind that in every action or
omission that we make in the classroom (and that is part of the hidden
curriculum), we are directly or indirectly affecting the
epistemological beliefs of our students. Something as simple as asking
our students what equation do we have to use to solve a particular
problem can foster (even unintentionally) an incorrect image of
science and how to learn physics, and ultimately affect the academic
achievement of our students.

\section*{Acknowledgments}
\label{sec:ack}

This work was carried out thanks to the financial support provided by
Agencia Nacional de Investigación e Innovación (National Agency for
Research and Innovation, Uruguay) and Consejo de Formación en
Educación (Council for Education Training, Uruguay) through the
project \textit{Conociendo e incidiendo sobre las concepciones
  epistemológicas de los futuros profesores de Física}
(FSED-3-2019-1-157320). We thank all the participants, especially the
teachers at Facultad de Ciencias (School of Science) who kindly agreed
to administer the CLASS questionnaire to their General Physics I
students.

\bibliography{/home/arturo/Dropbox/bibtex/mybib}

\bibliographystyle{unsrt}

\end{document}